# Competing zero-field Chern insulators in Superconducting Twisted Bilayer Graphene

Petr Stepanov[1*], Ming Xie[2], Takashi Taniguchi[3], Kenji Watanabe[3], Xiaobo Lu[1], Allan H. MacDonald[2], B. Andrei Bernevig[4] and Dmitri K. Efetov[1*]

1. ICFO - Institut de Ciencies Fotoniques, The Barcelona Institute of Science and Technology, Castelldefels, Barcelona, 08860, Spain
2. Department of Physics, University of Texas at Austin, Austin, TX 78712, USA
3. National Institute of Material Sciences, 1-1 Namiki, Tsukuba, 305-0044, Japan
4. Department of Physics, Princeton University, Princeton, New Jersey 08544, USA

*Correspondence to: petr.stepanov@icfo.eu, dmitri.efetov@icfo.eu

The discovery of magic angle twisted bilayer graphene (MATBG) has unveiled a rich variety of superconducting, magnetic and topologically nontrivial phases. The existence of all these phases in one material, and their tunability, has opened new pathways for the creation of unusual gate tunable junctions. However, the required conditions for their creation – gate induced transitions between phases in zero magnetic field – have so far not been achieved. Here, we report on the first experimental demonstration of a device that is both a zero-field Chern insulator and a superconductor. The Chern insulator occurs near moiré cell filling factor $\nu = +1$ in a hBN non-aligned MATBG device and manifests itself via an anomalous Hall effect. The insulator has Chern number $C = \pm 1$ and a relatively high Curie temperature of $T_c \approx 4.5$ K. Gate tuning away from this state exposes strong superconducting phases with critical temperatures of up to $T_c \approx 3.5$ K. In a perpendicular magnetic field above $B > 0.5$ T we observe a transition of the $\nu = +1$ Chern insulator from Chern number $C = \pm 1$ to $C = 3$, characterized by a quantized Hall plateau with $R_{yx} = h/3e^2$. These observations show that interaction-induced symmetry breaking in MATBG leads to zero-field ground states that include almost degenerate and closely competing Chern insulators, and that states with larger Chern numbers couple most strongly to the $B$-field. By providing the first demonstration of a system that allows gate-induced transitions between magnetic and superconducting phases, our observations mark a major milestone in the creation of a new generation of quantum electronics.

Recently discovered quantum phases in the flat-bands of $\theta_m \sim 1.1°$ magic angle twisted bilayer graphene (MATBG) include correlated insulators[1–5] (CI), superconductors[2,6–16] (SC), and interaction induced correlated Chern insulators[17–19, 35-38] (CCI). The CCIs can occur with different Chern numbers, and have $U(4)$ valley/spin ferromagnetism in the bulk and topologically protected states at device edges. The search for the exact nature of these exotic phases[20–23] and the competition[10,24–29] between them requires a complete understanding of the role of electronic interactions in the symmetry breaking of the non-interacting 4-fold spin and valley degenerate bands. The existence of multiple correlated phases in one materials platform opens up new possibilities for the creation of complex gate tunable junctions[30,31]. Among these the most interesting are junctions between superconducting and topological magnetic phases. The clean, gate defined homojunctions of these phases could pave new avenues for the creation of topological and spin-triplet superconductivity, as well as non-abelian particles, such as para-fermions and Majorana fermions[32]. However, the necessary requirements for such junctions, namely reversible gate tuning between SC and CCI states in a single device in zero magnetic field, has not been previously achieved.

The occurrences of CCI phases in MATBG at integer carrier filling per moiré unit cell $\nu$ is a result of electronic interactions that break the system's combined inversion and time-reversal symmetry $C_2T$. Breaking this symmetry can give rise to gapped valley polarized bands and to the formation of quantum Hall isospin ferromagnets[33–35] with a well-defined correspondence $(C, \nu)$ between Chern numbers $C$ and electron fillings $\nu$ [25,26]. However, while theory predicts the existence of a variety of competing CCIs in zero magnetic field[24,25], some of these states have so far only be observed at elevated perpendicular $B$-field, and hence above the critical field of the SC states. The experimental variability of the CCI state manifestation can be explained by the sensitivity of the competition between states to experimental parameters, such as strain and dielectric environment[10], and also possibly by complex networks of magnetic domain walls that can obscure the quantization of the CCIs. In hexagonal boron nitride (hBN) aligned MATBG devices with explicitly broken $C_2$ symmetry at the single particle level, CCIs with a quantized Hall conductance in zero magnetic field have been found at $\nu = +3$[18]. However, neither SC nor CIs were observed in these devices, demonstrating that the single-particle term in the Hamiltonian that favors one sublattice over the other in aligned hBN devices, alters the competition between states.

Here we report on the first observation of an interaction induced anomalous Hall effect (AHE) in hBN non-aligned MATBG. The CCI occurs near $\nu = +1$ filling and marks the formation of two (not yet fully quantized) $C = \pm 1$ correlated Chern insulators in zero magnetic field $B = 0$ T. This is the first observation of a CCI near $\nu = +1$ in any MATBG device, hBN aligned or non-aligned. The device also displays CIs with trivial gaps at $\nu = \pm 2$ and $\nu = +3$, as well as fully developed superconducting regions that allow direct gate-induced switching between SC and CCI states. Furthermore, in elevated out-of-plane $B$-field the device shows a pronounced sequence of perfectly quantized CCIs which follow the correspondence of $(C, \nu)$ of $(\pm 2, 0)$, $(\pm 4, 0)$, $(\pm 3, \pm 1)$, $(\pm 2, \pm 2)$ and $(\pm 1, \pm 3)$[36–41]. We find that these high field topological CCI states can both coexist with non-topological CI states and compete with the CCI states observed in the absence of magnetic field, in particular the $(\pm 1,1)$ and the $(3,1)$ states.

Fig. 1a shows the optical image of the four-terminal transport device, which consists of a graphite/hBN/MATBG/hBN heterostructure with a twist angle of $\theta = 1.08 \pm 0.01°$ ($n_s = 2.71 \times 10^{12}$ cm$^{-2}$). Crystallographic alignment between MATBG and hBN substrate layers is excluded by optical images of the naturally broken crystal edges of the individual layers, as well as by the absence of a charge neutrality gap and satellite resistance peaks in transport measurements (see SI for details). Here the longitudinal $R_{xx} = V_{xx}/I$ and Hall $R_{yx} = -V_{xy}/I$ resistance values are obtained from lock-in voltage measurements of $V_{xx}$ and $V_{xy}$, and the source-drain current $I$ (see Methods). The charge carrier density $n$ is capacitively controlled with a local back gate voltage $V_g$ on the graphite layer. We define the filling factor of carriers per moiré unit cell as $\nu = 4n/n_s$, where $n_s$ is the density of fully filled superlattice flat band and the prefactor 4 accounts for spin and valley degeneracy of each low-energy flat band.

Fig. 1b (top) shows the temperature and density dependent phase diagram of the device, via a color plot of $R_{xx}$ vs. $n$ and $T$ in zero magnetic field $B$, and Fig. 1b (bottom) shows the corresponding line-traces of $R_{xx}$ and $R_{yx}$ vs. $n$ at base temperature $T = 30$ mK. At integer filling factors of $\nu = \pm 2$ and $\nu = +3$ we observe strong, temperature activated $R_{xx}$ peaks, while $R_{yx}$ remains negligible, indicating the formation of correlated insulators with topologically trivial gaps, as reported in all previous studies[2,6,9]. In addition, the device shows dome-shaped superconducting regions in the $n$-$T$ phase space in which $R_{xx} = 0$ Ω (Fig. 1 b and c), and which have a critical temperature of up to $T_c \approx 3.5$ K (taken as 50% of normal states resistance at $\nu =$ -2.16). Differential resistance $dV_{xx}/dI$ measurements vs. bias current $I_{dc}$ and perpendicular magnetic field $B$ show characteristic diamond shapes (Fig. 1d), and Fraunhofer features (ED Fig. 8), with critical currents and critical magnetic field values of $I_c = 692$ nA and $B_c = 210$ mT. These results are overall in very good agreement with values found in previous reports on superconductivity in MATBG[2,6,9].

In stark contrast to all previous studies of MATBG, however, at a filling of $\nu = +1$, we observe a pronounced anomalous Hall effect (AHE) manifested by a non-zero Hall resistance which at $T = 50$ mK reaches a value close to the quantum of conductance $R_{yx} \approx 0.9h/e^2$, and goes hand-in-hand with a vanishingly small longitudinal resistance that approaches $R_{xx} = 0$ Ω (Fig. 1e). The sign of $R_{yx}$ can be flipped by applying a small perpendicular field $B < 200$mT, and shows a strong hysteresis loop between up and down sweeps of the field, which are centered around $B = 0$ T. The anomalous Hall effect occurs in a narrow density range from $\nu \approx$ +0.6 to $\nu \approx +1$. The strength of the hysteresis loop, which is defined as $\Delta R_{yx}/2 = (R_{yx}^{B\downarrow}-R_{yx}^{B\uparrow})/2$ ($B = 0$ T), has a maximum at a filling of $\nu \approx +0.84$ ($n = 0.57\times10^{12}$ cm$^{-2}$). $\Delta R_{yx}/2$ is quite robust at elevated temperatures, with an extracted Curie temperature of $T_c \approx 4.5$ K (Fig. 1f), and follows a thermally activated dependence with an extracted energy gap $\Delta \approx 2.41$ meV. Overall these findings are notably close to previous reports on magnetism in graphene moiré heterostructures[17,18,42–44], which have been interpreted as manifestations of an underdeveloped correlated Chern insulator with $|C| = 1$. The absence of quantization at the present time may be due to quasiparticle delocalization, possibly assisted by transport pathways along domain walls.

While at $B = 0$ T we only observe two CCI states with indices (±1, 1), we see many more CCI states at high $B$-fields. The high $B$-field phase diagrams in Fig. 2a and b shows $R_{xx}$ and $R_{yx}$ measurements as a function of $n$ and $B$ at $T = 50$ mK, and Fig. 2c displays the schematic of the most dominant features of the resulting Landau Fan diagram. It consists of a multitude of regions with quantized Hall conductance $R_{yx} \sim h/Ce$ and $R_{xx} \sim 0$ Ω, which form wedge-like areas. These states follow a linear slope in the $n$-$B$ phase space, which is defined by the Streda formula[45] $dn/dB = Ce/h$, and originate from different fillings $\nu$ at $B = 0$ T. In addition, we also see highly resistive CIs at fillings $\nu = \pm 2$ and $\nu = +3$, which due to their topologically trivial gaps with Chern number $C = 0$, form vertical regions in the $n$-$B$ phase space. While overall, we see a multitude of Landau level-like gaps, we find a clear hierarchy of gaps where the states with the corresponding indices of $(C, \nu)$ are particularly strongly pronounced - (±2, 0), (±4, 0), (±3, ±1), (±2, ±2) and (±1, ±3). These states are quantized at much lower fields and have almost an order of magnitude larger extracted gaps than typical LLs in the system, and have been recently interpreted as interaction driven CCIs, which are stabilized by a small $B$-field[35].

We compare the experimental findings with the theoretical phase diagram of MATBG[22–25,46], which predicts a series of CIs and CCIs (Fig. 3). While the finer details of this phase diagram (in particular valley-spin polarizations) are not completely settled, the competition between states with different Chern numbers, if not their energetic order, is similar in most theoretical scenarios. In order to motivate the theoretical phase diagram, we consider two limits: first, the flat-band limit, where the kinetic energy of the bands is artificially tuned to zero, and second, the chiral limit, where the AA-region hopping between the MATBG bilayers is artificially tuned to zero (the AB-region coupling is still considered). In this "chiral-flat" limit, the system enjoys a large $U(4) \times U(4)$ symmetry, which allows for the exact determination of the ground-states of the Coulomb interaction Hamiltonian. Due to the single particle topology of MATBG, the two flat bands at each valley and spin bands can be labeled by a Chern number $C = 1$ and $C = -1$, related by $C_2$ symmetry[22–24,46–49]. In the chiral flat limit, at each filling $\nu$, the interaction favors the successive occupation of such Chern bands – reminiscent of quantum Hall ferromagnetism, leading to many-body degenerate ground states (which can be classified by $U(4) \times U(4)$ representations) at filling $\nu$ with Chern number $(4 - |\nu|)$, $(2 - |\nu|)$, ..., $(|\nu| - 2)$, $(|\nu| - 4)$. We interpret these "chiral-flat" limit states as the low-energy states which compete against each other in the realistic system and which we observe in this study.

Realistic MATBG is however, not exactly in the chiral-flat limit. Perturbation away from the chiral limit (but still not considering the kinetic energy of the bands) provides insight into the nature of the correlated states: the lowest Chern number correlated states win, while the larger Chern number correlated states acquire a finite energy above the ground-state. Hence at filling $\nu = 0, \pm 2$ the theoretical ground states have Chern number 0, while at $\nu = \pm 1, \pm 3$ has theoretical ground states have Chern number $|C| = 1$, all of which are now ferromagnetic in a lower $U(4)$ symmetry that still exists in the flat-band limit. A magnetic field lowers the energy of the large Chern number states with respect to those of the low Chern number states. At filling $\nu = \pm 1$ a first order phase transition between the low-field $|C| = 1$ ground-state and the high-field $|C| = 3$ ground-state is predicted to occur at $B_2 = 0.5$ T[25]. At filling $\nu = \pm 2$ a first order phase transition between the low-field $|C| = 0$ ground-state and the high-field $|C| = 2$ ground-state is predicted to occur at $B_1 = 0.2$ T, although these numbers should only be indicative of order-of magnitude.

Perturbation theory predicts the polarization of the Chern correlated states upon introduction of kinetic energy, when the symmetry of the system lowers even further to $U(2)$ spin-charge rotation per valley. The Chern number $C = 0$ and $0 < |C| < 4 - |\nu|$ states at integer fillings $\nu$ are fully and partially intervalley coherent respectively, while the states with Chern number $|C| = 4 - |\nu|$ are valley polarized. These results are perturbative away from the chiral-flat limit. The theoretical predictions are only as good as the region of validity of the perturbation allows. Numerical results based on exact diagonalization and DMRG[23,46,50] suggest that some of the Chern insulating ground-states, particularly at $\nu = \pm 3$, do not survive for realistic MATBG parameters, and that a competition occurs between (nematic) metal, momentum $M(\pi)$ stripe, and $K$-CDW orders and metallic states with no broken symmetries.

We now analyze the correspondence between the experimental and the theoretical phase diagrams. At $\nu = \pm 2$, the experiment and theory are in agreement with a zero field Chern number $C = 0$ state and an in-field $|C| = 2$ state. In zero field at $\nu = +3$, the experimental Chern number is $C = 0$, while for $\nu = -3$ we do not find any experimental signatures of neither, the CI nor a CCI. These zero field states conflict with the perturbation-derived $|C| = 1$ states, however at high $B$-field these transition to $|C| = 1$ states, as predicted by theory. We find that the theoretically predicted low field $|C| = 1$ state is the only possible option arising from a translationally invariant interaction driven Chern bands at at $\nu = \pm 3$, although $C = 0$ insulators are allowed if mixing with remote bands plays a role[24]. Its absence could suggests the presence of translational symmetry-broken states at this filling, which is further supported by numerical results[25]. At $\nu = +1$, the experimental discovery, presented here of a $|C| = 1$ is in tune with the theoretically predicted ground-state. Most strikingly, the in-field transition from the $|C| = 1$ to the higher, $|C| = 3$ state also corresponds to the theoretically predicted state. Such agreement at $\nu = +1$ suggests that further samples can reveal similar physics at $\nu = -1$.

We examine the observed $B$-field induced transitions of the phases originating from the $\nu = +1$ filling in detail in Fig. 4b, which shows a color map of $R_{yx}$ vs. $n$ and $B$. At low fields $B < 0.5$ T the phase diagram is mainly defined by the AHE hysteresis loop on the hole doped side of $\nu < +1$. The center of the hysteresis loop shifts in $B$-field in agreement with the Streda formula for a Chern number of $C = -1$ for positive and $C = +1$ for negative $B$-field (Fig. 4a). The sign of the Chern numbers is consistent with the sign of $R_{yx}$ for the AHE, which maintains positive values in positive $B$, while the normal Hall effect at higher $n$ produces a negative $R_{yx}$. While $\Delta R_{yx}$ is strongest around $\nu \approx +0.84$, close to $\nu = +1$ it is strongly suppressed and its coercive $B$-field values are increased (Fig. 4d and ED Fig. 4). Sharply at $\nu > +1$ we observe a sign reversal of $\Delta R_{yx}$, which however becomes much weaker. Changes in the sense of the hysteresis loop and the magnitude of the hysteresis loop across the gap are expected to be a common feature of Chern insulators[43,51] (See SI and Fig. 4e). These low field CCI states gradually disappear above $B > 0.5$ T, and we observe the onset of an CCI which also follows the Streda formula with Chern number of $C = +3$ and shows well quantized Hall plateaus $R_{yx} = h/3e^2$ ($R_{xx} = 0$ Ω) shown in Fig. 4c. The absence of a $C = -3$ state is likely due to competing metallic states with no broken symmetries as explained in the supplementary material. These findings clearly establish that just like theory predicts, the ground states of MATBG consist of closely competing CCIs, for which $B$-field can act as a tuning knob, which couples strongest to the states with higher Chern number.

To summarize - our findings shed new light on the underlying ground states of MATBG, and show that even in zero-field these form nearly degenerate and competing interaction-induced Chern insulators, which can be further tuned by weak magnetic fields. The exact sequence of these phases in zero-field and their evolution in $B$-field gives detailed information about the competition of these phases and allows to understand the exact microscopic mechanism that drive their formation, through comparisons to ongoing theoretical models. The zero-field coexistence and gate tunability of these magnetically and topological nontrivial phases with superconducting phases, presents a remarkable opportunity to electronically hybridize these phases through engineering of complex gate induced junctions, and will lead to the creation of ever more complex quantum phases based on the MATBG platform.

Acknowledgements:
We are grateful for fruitful discussions with Frank H. L. Koppens and Ashvin Vishwanath. D.K.E. acknowledges support from the Ministry of Economy and Competitiveness of Spain through the "Severo Ochoa" program for Centres of Excellence in R&D (SE5-0522), Fundació Privada Cellex, Fundació Privada Mir-Puig, the Generalitat de Catalunya through the CERCA program, funding from the European Research Council (ERC) under the European Union's Horizon 2020 research and innovation programme (grant agreement No. 852927)" and the La Caixa Foundation. B.A.B. was supported by the DOE Grant No. DE-SC0016239, the Schmidt Fund for Innovative Research, Simons Investigator Grant No. 404513, the Packard Foundation, the Gordon and Betty Moore Foundation through Grant No. GBMF8685 towards the Princeton theory program, and a Guggenheim Fellowship from the John Simon Guggenheim Memorial Foundation. Further support was provided by the NSF-EAGER No. DMR 1643312, NSF-MRSEC No. DMR-1420541 and DMR-2011750, ONR No. N00014-20-1-2303, Gordon and Betty Moore Foundation through Grant GBMF8685 towards the Princeton theory program, BSF Israel US foundation No. 2018226, and the Princeton Global Network Funds. P.S. acknowledges support from the European Union's Horizon 2020 research and innovation programme under the Marie Skłodowska-Curie grant agreement No. 754510. A.H.M and M.X were supported by DOE grant DE- FG02-02ER45958 and Welch Foundation grant F1473.

Author contributions:
D.K.E and P.S. conceived and designed the experiments; P.S. fabricated the devices and performed the measurements; P.S., D.K.E., B. A. B., M.X. and A.H.M. analyzed the data; B.A.B, M.X. and A.H.M performed the theoretical modeling; T.T. and K.W. contributed materials; D.K.E. and X. L. supported the experiments: P.S., D.K.E., B. A. B, M.X. and A.H.M wrote the paper.

**Supplementary Information** is available for this paper.

**Correspondence and requests for materials** should be addressed to P.S. and D.K.E.

**Competing interest.** Authors claim no competing interest.

**Methods and materials.**
Device fabrication.
Our devices are fabricated using a "cut-and-stack" dry transfer method (ED Fig. 1). We assemble our heterostructure from top to bottom using a sacrificial poly-bis-phenol A carbonate (PC) layer placed on top of a PDMS stamp (polydimethylsiloxane). All flakes used for the assembly are prepared by mechanical exfoliation process on $Si^{++}$/$SiO_2$ chips. First, we pick up a top hBN flake, followed by a subsequent pick up of the first graphene half, which was precut in two pieces using an atomic force microscopy tip. During the next step, the transfer stage is rotated by ~1.1° introducing an interlayer twist between graphene layers, followed by a second graphene pickup process. The partial hBN/MATBG stack is then used to pick up a bottom hBN flake (~7nm), followed by a pick-up of a graphite stripe (~3nm thick) that is used as a high-quality local back gate. At the next steps, the quadruple stack hBN/MATBG/hBN/graphite is etched into a multi-terminal Hall bar geometry and coupled to 1D metallic contacts (Cr/Au 5/40 nm).

Measurements.
Transport measurement were performed in a cryogen-free dilution refrigerator using lock-in amplifiers with low AC excitation current in the range 1-5 nA at the excitation frequency of 19.111 Hz. The direct current measurements shown in ED Fig. 9 were performed using a SR560 DC voltage preamplifier in combination with a Keithley 2700 multimeter. Most of the data shown in this study is taken at base temperatures of the dilution refrigerator 30-50mK and available magnetic fields up to 8 T.

Twist angle extraction.
To extract the twist angle we analyze the phase diagram shown on the ED Fig. 3. Using the relation $n_s = 8\theta^2/\sqrt{3}a^2$, where $\theta$ is the interlayer twist angle, $n_s$ is the charge carrier density corresponding to the fully filled superlattice unit cell and $a$=0.246 nm is the interatom distance in single layer graphene, we determine the twist angle in our device $\theta$=1.08±0.01°. $n_s$ is derived from the quantum oscillations emanating outside of the fully filled flat bands.

**Figures.**

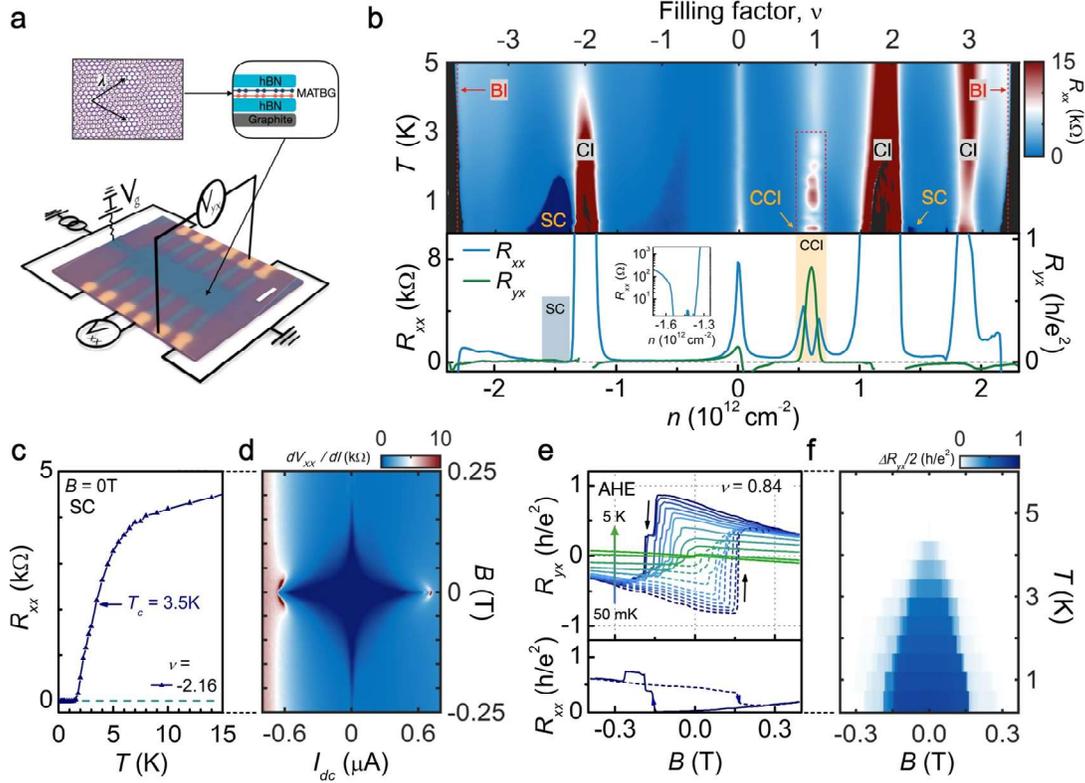

Figure 1. | **Coexistence of magnetism and superconductivity in MATBG. (a)** Optical image and experimental transport measurements setup of the locally gated MATBG device. **(b)** Upper panel. Colormap $R_{xx}$ vs. $n$ and $T$. A multitude of emergent strongly correlated phases including SCs, CIs and CCI states coexisting in the same phase space. Dark blue regions indicate SC phases with $R_{xx}$ values below the noise level. Black regions indicate data points with quenched drain current. BI denotes band insulators. Lower panel shows $R_{xx}$ and $R_{yx}$ vs. $n$ line traces at 20 mT. The data points, for which drain current quenches, were dropped out. The sample is tuned by the back gate between AHE and SC phases. The inset demonstrates a zoom-in image around the SC pocket with vanishing longitudinal resistance values. **(c)** Temperature dependence $R_{xx}$ vs. $T$ of the superconducting domain at optimal doping $\nu$=-2.16 shows critical temperature $T_c$ = 3.5 K (defined by the temperature of 50% of the normal metal state resistance). **(d)** Differential resistance $dV_{xx}/dI$ vs. direct current bias $I_{dc}$ and magnetic field $B$. A "diamond" - like feature corresponds to SC phase with optimal doping at $\nu$=-2.16. **(e)** Upper panel. Hall resistance $R_{yx}$ vs. $B$ taken for increasing and decreasing magnetic fields. Hysteresis loops are indicative of the incipient Chern insulator $|C|$=1 at filling $\nu$=+0.84. Black arrows show the $B$-field sweep directions. Lower panel shows $R_{xx}$ vs. $B$ line trace taken at $T$=50 mK. **(f)** AHE resistance $\Delta R_{yx}/2$ vs. $B$ and $T$ taken at $\nu$=+0.84. $\Delta R_{yx}/2$ is defined by subtracting $R_{yx}(B)$ as magnetic field decreases from $R_{yx}(B)$ as $B$ increases $\Delta R_{yx}/2=(R_{yx}^{B\downarrow}-R_{yx}^{B\uparrow})/2$. The hysteresis disappears above $T_C$~4.5 K.

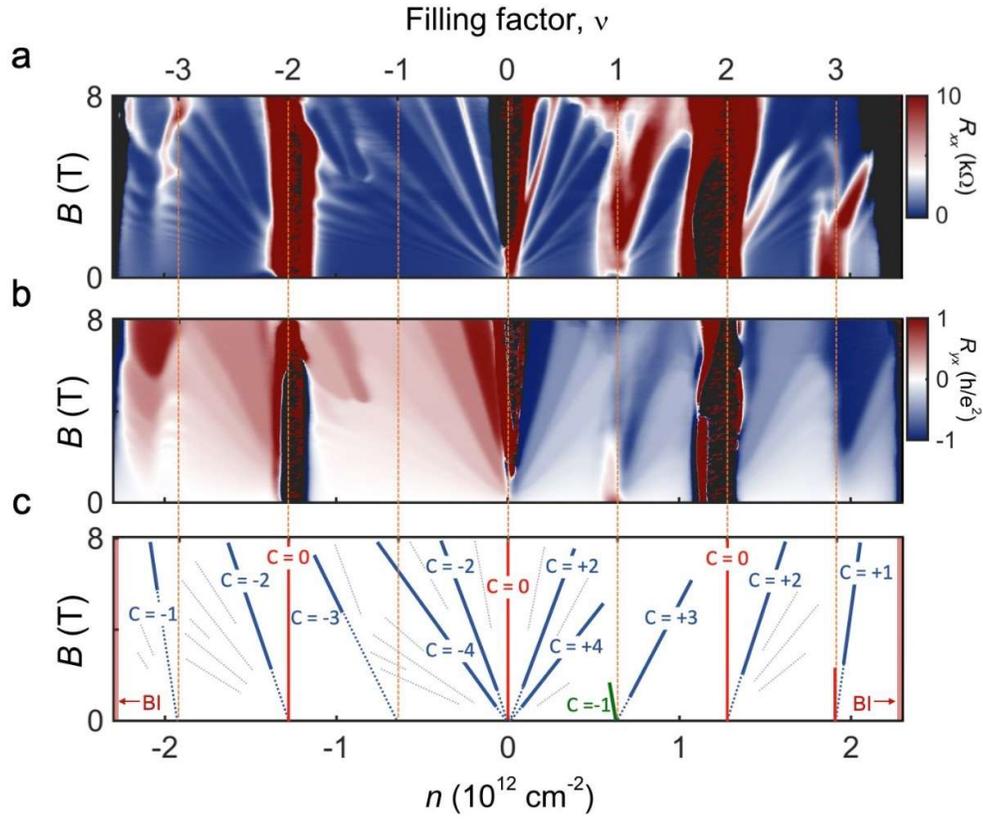

Figure 2. | **Chern insulators in MATBG at T=50 mK. (a)-(b)** $R_{xx}$ and $R_{yx}$ vs $n$ and $B$ over the full range of charge carrier densities of the low-energy flat bands. The phase space exhibits a zero-field Chern number -1 insulator (green solid line) and a set of high-field Chern insulators carrying numbers ±1, ±2, and ±3 (solid blue lines). We observe the full set of Chern insulators emanating from partial fillings of the superlattice unit cell ($\nu$, $C$) = (±1, ±3), (±2, ±2), (±3, ±1). **(c)** Schematic image, illustrating the experimental phase diagram shown in (a)-(b). The solid lines correspond to well developed Chern insulators with Hall conductance quantized at integer number of $e^2/h$ in the broad regions of the phase space in (b). Dashed lines show other quantum Hall states.

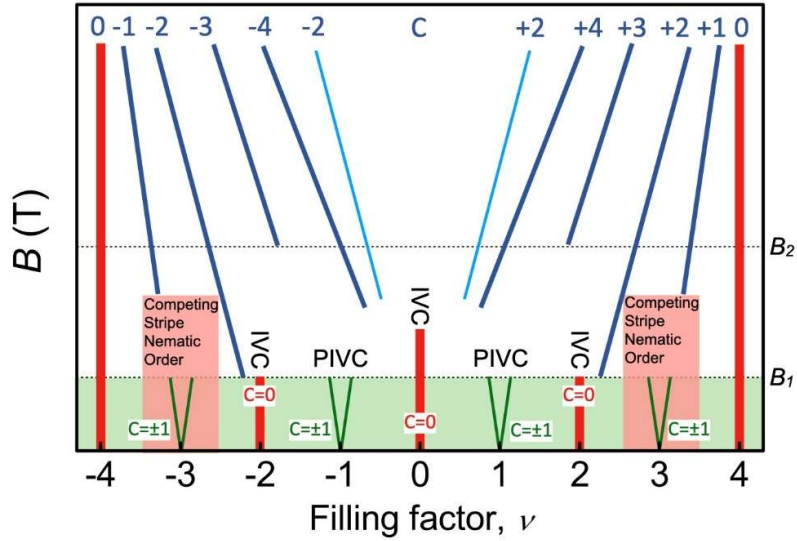

Figure 3. | **Phase diagram of MATBG in a magnetic field**. Large magnetic field favors the largest Chern number possible many-body states given the single-particle MATBG bands. These states exhibit the maximum valley polarization allowed by the filling. At low fields, the system undergoes phase transitions from the high-field ground-states, and the theoretical situation is more complex. While at filling 0 and ±2, the many-body states are correlated insulators with zero Chern number and exhibit inter-valley coherence, at filling ±3, there is strong competition between a correlated $C=\pm 1$ Chern insulator and several $C=0$ translational or rotational symmetry broken states. Experimentally, no CCI is obtained at this filling, indicating that the CDW or nematic states win. At filling ±2, the ground-state switches from $|C|=2$ in high field to $C=0$ at low field, in agreement with experiment. Nontrivially, at filling ±1, the ground-state switches from $|C|=3$ in high field to $|C|=1$ in low field, in agreement with our new experimental results. Dark blue lines above $B_1$ show valley polarized CCI states. PIVC denotes partially inter-valley coherent and IVC – inter-valley coherent states.

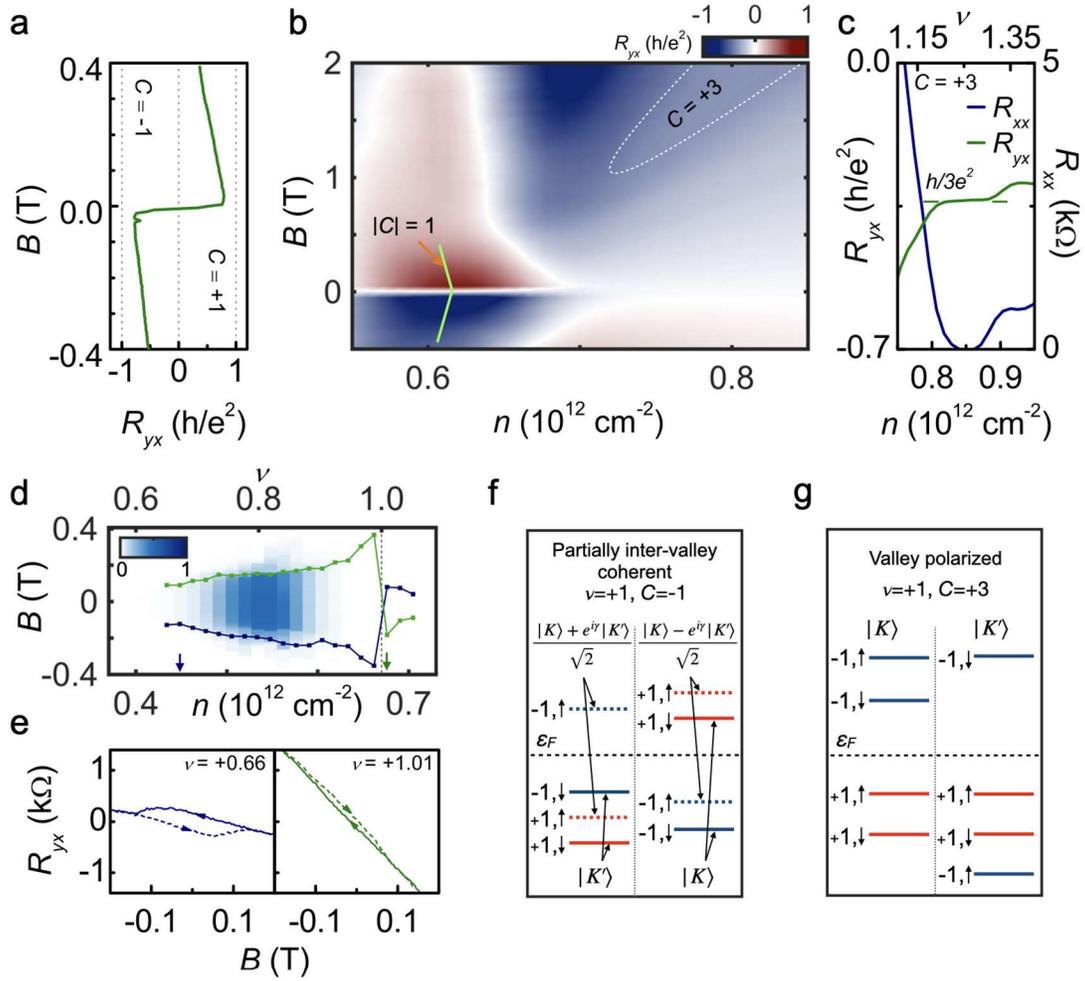

Figure 4. | **Competing zero-field Chern insulators in MATBG at $T$=50 mK.** (a) $R_{yx}$ vs. $B$ trace taken along the green line shown in (b). The slope of $R_{yx}$ minimizes when the line is chosen along $dn/dB = \pm e/h$. (b) $R_{yx}$ vs. $n$ and $B$ close to $\nu$=+1. The phase space exhibits zero-field Chern number 1 CCI (green line) and high-field Chern number 3 CCI (outlined by the white dashed line). (c) $R_{yx}$ and $R_{xx}$ vs $n$ taken at $B$=2.5 T exhibits a single quantum Hall-like fully quantized $h/3e^2$ plateau indicative of a CCI with Chern number 3. (d) AHE resistance $\Delta R_{yx}/2$ vs. $n$ and $B$ taken from the pair of contacts adjacent to one in Fig. 1e. Green and blue symbols show coercive field values. Note the switch of the magnetization sign upon crossing $\nu$=+1 (dashed line). Colorscale is set to von Klitzing constant $h/e^2$. (e) $R_{yx}$ vs. $B$ line traces taken at charge carrier densities color-coded to arrows shown in (d). (f)-(g) Schematics of the $C$=-1 and $C$=+3 states at filling $\nu$=+1. Per graphene valley, the two bands of MATBG can be transformed into two bands one with Chern number 1, the other with Chern number -1. The ground-states of the interacting Hamiltonian (exact in some limit) then correspond to specific ways of filling these states. For $C$=-1, we find that a partially valley coherent state (dashed lines, basis formed by wavefunctions from the two valleys) is preferred, while for $C$=+3 (the high-field ground-state), the largest valley polarization state wins.

# Supplementary Information: Competing zero-field Chern insulators in Superconducting Twisted Bilayer Graphene

Petr Stepanov[1*], Ming Xie[2], Takashi Taniguchi[3], Kenji Watanabe[3], Xiaobo Lu[1], Allan H. MacDonald[2], B. Andrei Bernevig[4] and Dmitri K. Efetov[1*]

1. ICFO - Institut de Ciencies Fotoniques, The Barcelona Institute of Science and Technology, Castelldefels, Barcelona, 08860, Spain
2. Department of Physics, University of Texas at Austin, Austin, TX 78712, USA
3. National Institute of Material Sciences, 1-1 Namiki, Tsukuba, 305-0044, Japan
4. Department of Physics, Princeton University, Princeton, New Jersey 08544, USA

*Correspondence to: petr.stepanov@icfo.eu, dmitri.efetov@icfo.eu

**A. Check for alignment to hBN.**

To ensure that there is no alignment of crystallographic edges between graphene and either encapsulating hBN, we study a relative twist angle mismatch between the layers under optical microscope (Extended Data Fig. 1). White dashed line in the Extended Data Fig. 1a shows a naturally broken graphene edge that we ascribe to either zig-zag or armchair type. Similarly, orange and blue dashed lines in the Extended Data Fig. 1b and c, respectively, show naturally broken edges of the top and bottom hBN layers. White, orange and blue dashed lines in Extended Data Fig. 1d correspond to ones shown in the Extended Data Fig. 1a-c, and the numbers indicate relative twist angles between the naturally broken graphene edge and those found in hBN layers.

Extended Data Fig. 2 demonstrates Arrhenius plots for the superlattice unit cell fillings $\nu=0,\pm2,\pm4$. CNP at $\nu=0$ shows a very weak dependence upon the decreasing $T$, which suggests no thermally activated gap. This is in a stark contrast with the hBN-aligned devices, which usually show large CNP gaps $\Delta_0 \approx 6-7$ meV due to sublattice symmetry breaking. CI states at $\nu=\pm2$ show strong insulator-like thermal activation behavior with $\Delta_{-2}=1.49$ meV, $\Delta_{+2}=1.67$ meV. We also extract band insulator gaps $\Delta_{-4}=32$ meV, $\Delta_{+4}=45$ meV. We note that the band structure exhibits strong particle-hole asymmetry with the stronger band insulator on the conduction flat band side.

Lastly, Extended Data Fig. 3 demonstrates an extended range phase diagrams $R_{xx}$ as a function of $n$ and $T$ beyond the flat band region. We note, that we do not observe additional satellite resistance peaks due to the interplay between the hBN/graphene moiré superlattice and the magnetic length that could be possibly assigned to alignment to hBN substrate. We note that in case of perfect (0°) alignment between hBN and graphene we expect to observe additional resistance peaks around $n = 2.4\times10^{12}$ cm$^{-2}$, which are completely absent in our measurements both inside and outside the flat band region.

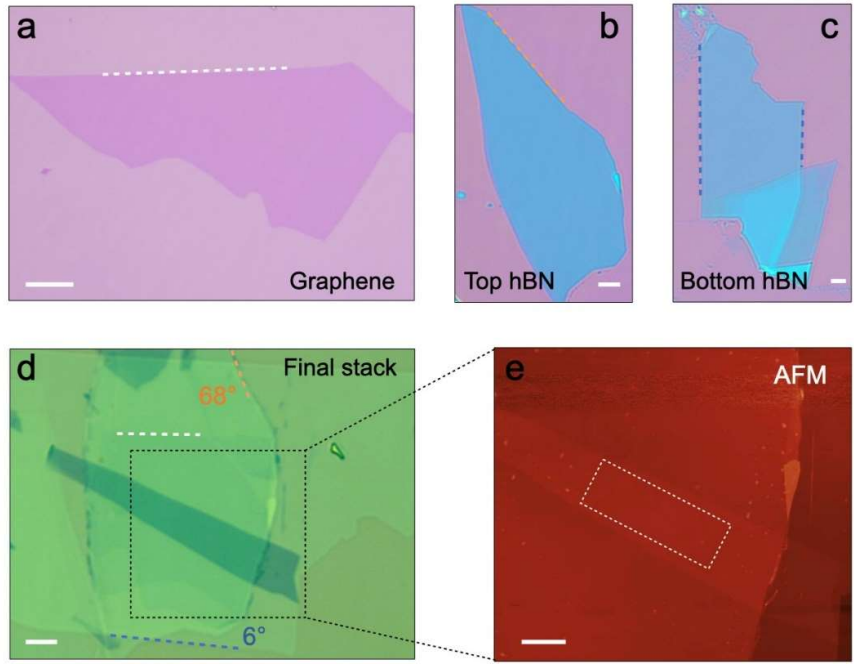

Extended Data Figure 1. | **Alignment to hBN. (a) - (c)** Optical micrographs of monolayer graphene (a), top (b) and bottom hBN (c) used for stacking the heterostructure shown in (d). **(d)** Optical image of the final stack. The numbers show relative misalignment angles between the white line (graphene edge) and bottom (blue) and top (orange) hBN edges. **(e)** Atomic force microscopy image of the stack shown in (d). The white dashed line box shows the device area. White bars correspond to 5 μm.

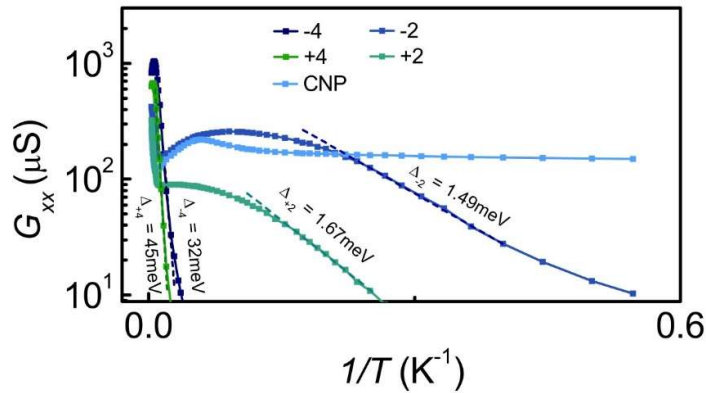

Extended Data Figure 2. | **Arrhenius plots for integer fillings $\nu=0, \pm2, \pm4$.** Extracted gaps are $\Delta_{-2}$=1.49 meV, $\Delta_{+2}$=1.67 meV, $\Delta_{-4}$=32 meV and $\Delta_{+4}$=45 meV. CNP resistance demonstrates a weak dependence on $T$ suggesting no thermally activated gap thus pointing towards no explicit sublattice symmetry breaking due to alignment to hBN.

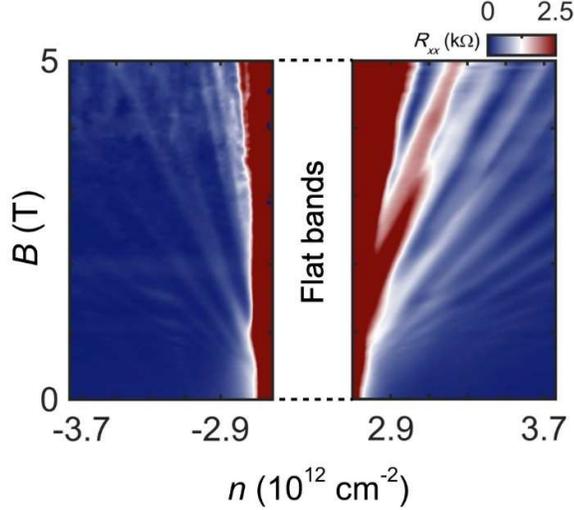

Extended Data Figure 3. | $R_{xx}$ **vs.** $n$ **and** $T$ **outside the flat band region.** The data demonstrates an absence of the satellite peaks that might be caused by a crystallographic alignment to hBN.

**B. Extended data on AHE close to $\nu=+1$ in MATBG.**

In this section we show additional measurements of the AHE state close to $\nu=+1$. We plot hysteresis loops taken at different charge carrier densities in Extended Data Fig. 4a for the same pair of contacts as in Fig. 1e in the main text (D-H in Extended Data Figure 7). In Extended Data Fig. 4b we show the evolution of anomalous Hall resistance $\Delta R_{yx}/2$ as a function of $n$ and $B$ extracted from a set of hysteresis loops close to $\nu=+1$. Maximized AHE resistance ($\Delta R_{yx}/2$) densely localizes close to $\nu \approx +0.84$.

Extended Data Fig. 4c demonstrates $\Delta R_{yx}/2$ in units of $h/e^2$ as a function of $n$ taken along the $B=0$ T line in Extended Data Fig. 4b. Upon increasing the superlattice filling from $\nu \approx +0.84$ to $\nu=+1$ we observe a strong suppression of the AHE resistance, while the coercive field values keep increasing for both negative and positive $B$ (blue and green data points in Extended Data Fig. 4b, respectively) indicative of a gradually suppressed coupling to the external field. In fact, we observe that the coercive field values maximize at $\nu \approx +0.98$ very close to the full integer filling. Upon closer inspection we find that this feature is likely associated with a weak magnetization reversal when the superlattice filling changes from the hole- to electron-doped side of $\nu=+1$ similar to the previously reported pattern of quantized AHE at $\nu=+3$ in hBN-aligned MATBG and twisted mono-bilayer graphene (see Extended Data Fig. 5).

We find this observation consistent with the divergence of the coercive field while the sample doping reaches exactly $\nu=+1$. For a fixed valley polarization the total magnetization can change sign when passing through zero, therefore, being discontinuous when the chemical potential crosses the gap; the coupling to the magnetic field vanishes while approaching to $\nu=+1$, thus, diverging the coercive field. In case of small Chern numbers (as in case of MATBG) the input of the bulk component of magnetization may be dominant resulting in a less robust magnetic states due to the higher influence of disorder. Ideally, in case of a minimized bulk magnetization, the edge magnetization would allow for the reversibility of the CCI at $\nu=+1$. However, without alignment to hBN the probability to observe a fully reversible CCI is very low due to the absence of the exchange splitting between the Fermi level and the remote bands.

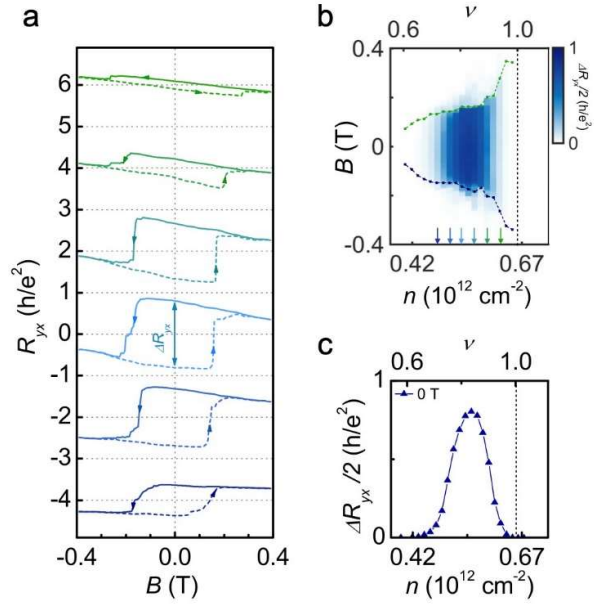

Extended Data Figure 4. | **Anomalous Hall effect at $\nu=+1$ without alignment to hBN.** (a) A set of hysteresis loops taken at superlattice density fillings $\nu$=0.74, 0.78, 0.82, 0.86, 0.90 and 0.94 (from bottom to top) and shifted by $2h/e^2$ for clarity. The data are obtained from the same pair of contacts as shown in Fig. 1e (D-H in Extended Data Figure 7). $T$=50 mK. (b) AHE resistance $\Delta R_{yx}/2$ as function of $n$ and $B$. Green and blue dashed lines with symbols show coercive field values for positive and negative $B$, respectively. Colorscale is set to von Klitzing constant $h/e^2$. Arrow colors indicate hysteresis loops shown in (a). (c) Line trace taken along $B$=0 T shows maximized AHE resistance around $\nu=+0.84$.

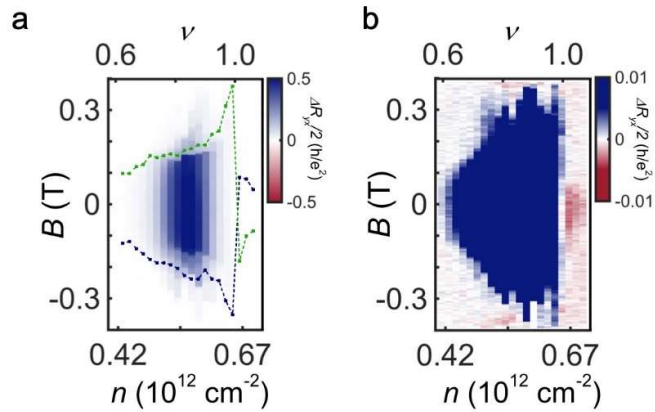

Extended Data Figure 5. | **Signatures of weak magnetization reversal at $\nu=+1$.** (a) AHE resistance $\Delta R_{yx}/2$ as function of $n$ and $B$ taken from a pair of contacts C-I (see Extended Data Fig. 7), the same dataset reported in Fig. 4d on a different colorscale. (b) Same as (a) shown on a significantly enhanced colorscale reveals faint magnetization reversal signatures for charge carrier densities $\nu=+1+\delta$. Note an abrupt anomalous Hall resistance sign change from strongly negative to weakly positive upon crossing over $\nu=+1$.

## C. Correlated Chern insulators stabilized by magnetic field.

By applying higher magnetic field, we reveal a sequence of the quantum Hall-like plateaus that we assign to the high-field correlated Chern insulators. The line cuts shown in Extended Data Fig. 6 demonstrate the dominant sequence of the plateaus that match exceptionally well with the theory of the predicted CCIs at high magnetic fields above $B_1$ (Fig. 3 in the main text).

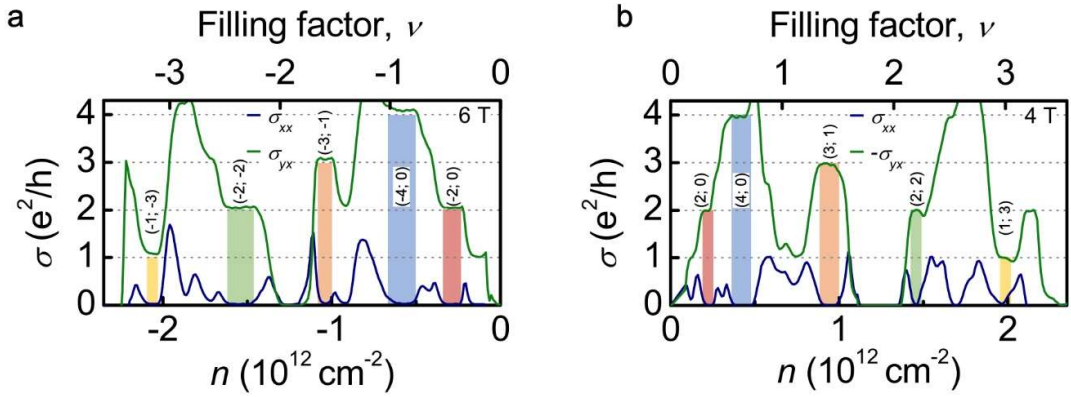

Extended Data Figure 6. | **Longitudinal and Hall conductance at high magnetic field.** Hall conductance $\sigma_{yx}$ vs. $n$ at **(a)** 6T and the valence flat band and **(b)** 4T and the conductance flat band. The sequence reveals a set of strongly quantized $\sigma_{yx}$ with $(C, \nu) = (\pm 2, 0), (\pm 4, 0), (\pm 3, \pm 1), (\pm 2, \pm 2)$ and $(\pm 1, \pm 3)$.

## D. Angle homogeneity in the studied sample.

We perform two terminal conductance measurements between pairs of adjacent contacts to mesoscopically probe twist angle homogeneity. All pairs exhibit a presence of strong correlations visible by the emergence of CI conductivity minima at $\nu=+1, \pm 2, +3$. This indicates a very low twist angle disorder, which can also mediate the observation of AHE state at $\nu=+1$.

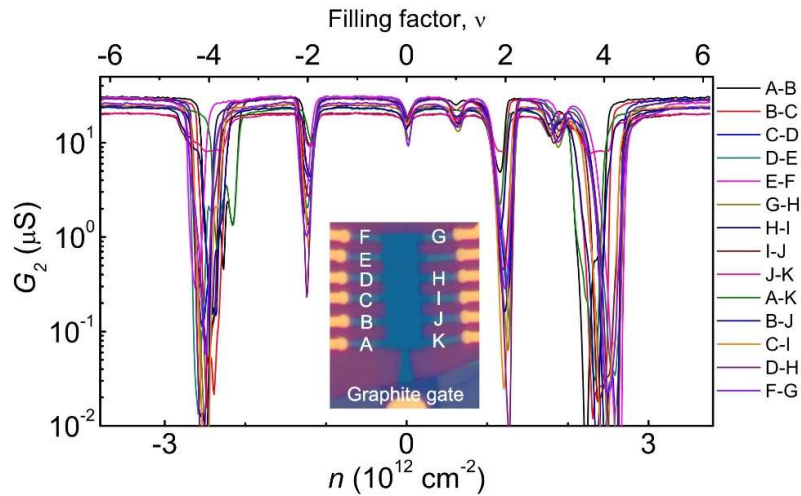

Extended Data Figure 7. | **Two-terminal conductance data.** Line traces correspond to the two terminal conductance $G_2$ vs. $n$ taken for all available pairs of contacts. The legend indicates corresponding pairs.

We quantify the twist angle disorder by extracting twist angles relative to the pair of contacts reported in the main text (C-D in Extended Data Figure 7). First, we find the deviation of the charge carrier density corresponding to the $\nu = \pm 2$ for all pair of contacts. Second, we compare this value to the one extracted from the SdH oscillations for the pair C-D (see Methods). Last, we extract a conformed twist angle using the formula $n_s = 8\theta^2/\sqrt{3}a^2$. As a result, we find the deviations of the charge carrier density corresponding to the fully filled superlattice, which fall in the range $n_s = (2.71\pm0.04)\times10^{12}$ cm$^{-2}$. Thus, the absolute twist angle variation across the sample is $\theta = 1.08\pm0.01°$.

### E. Hall density data.

We extract the Hall density from the low-field Hall resistance data using the relation $n_H = -B/eR_{xy}^{antisym}$, where $R_{xy}^{antisym} = (R_{xy}(B) - R_{xy}(-B))/2$ is an antisymmetric component of the measured Hall resistance used to eliminate any symmetric-in-$B$ input. The Hall density experiences a few clear resets along the charge carrier line indicative of new Fermi surface formations at $\nu = +1, \pm 2, \pm 3$. This data is consistent with high magnetic field data shown in Fig. 2 in the main text, where we observe new sets of well-quantized QH-like plateaus emanating from each of these integer fillings. Note the switch of Hall density sign inside the AHE region denoted by the black dashed lines.

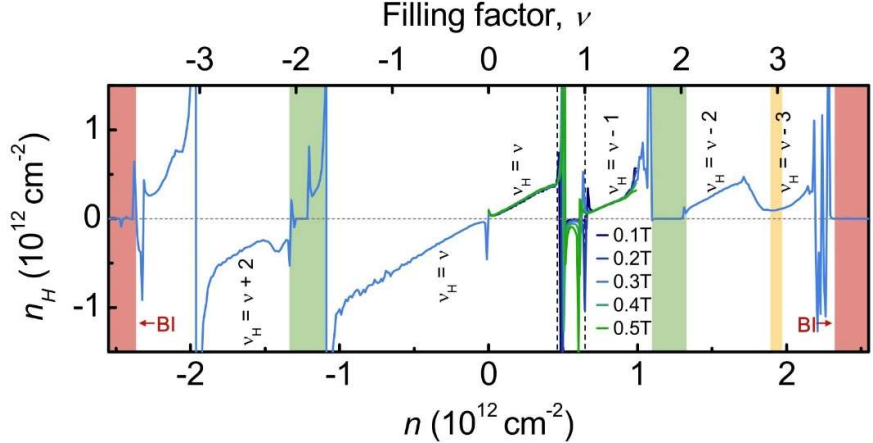

Extended Data Figure 8. | **Hall density measurements.** The light-blue line trace shows $n_H$ vs. $n$ taken at 0.3 T. The light green and light yellow stripes show the position of CI states, at which we also observe clear signatures of Hall density resets. Black dashed lines mark the region, where we observe signatures of AHE. Interestingly, the Hall density changes sign inside and outside this region. In addition, we plot Hall density $n_H$ vs. $n$ close to $\nu = +1$ for other magnetic field values (0.1, 0.2, 0.4, 0.5 T).

## F. Additional data on superconductivity and magnetic hysteresis.

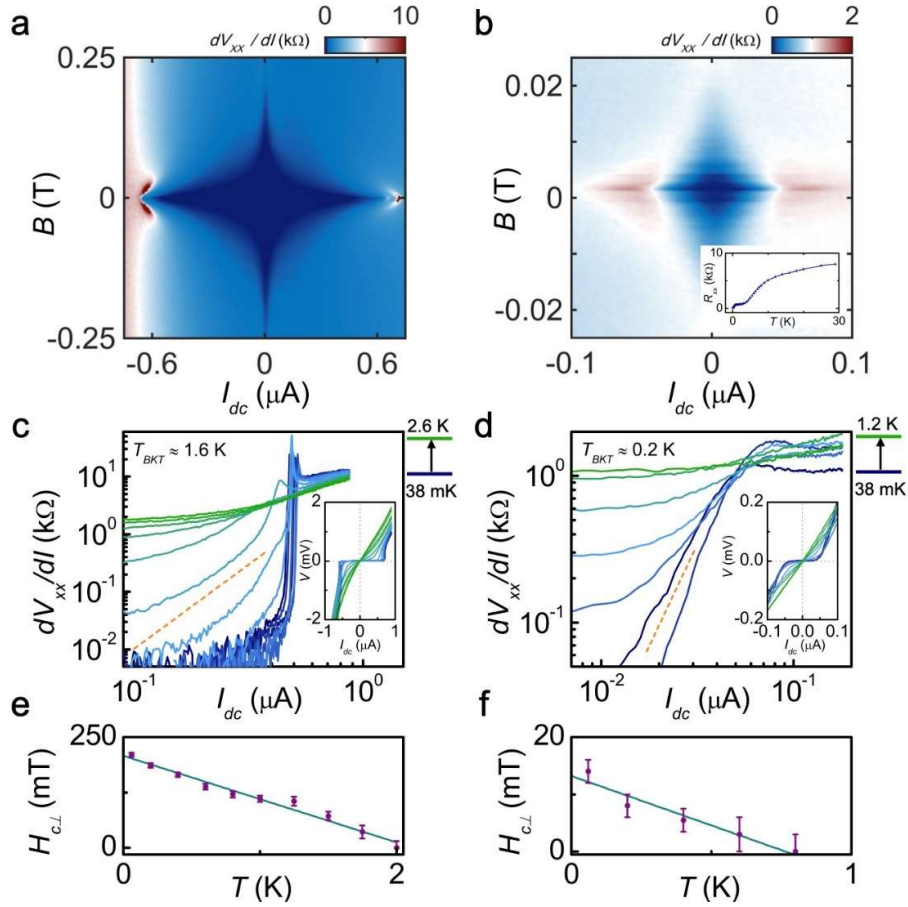

Extended Data Figure 9. | **Full characterization of SC phases.** The left- (right-) hand side of the figure corresponds to SC pockets with optimal doping $\nu=-2.16$ ($\nu=2.10$). **(a)-(b)** Differential resistance $dV_{xx}/dI$ as a function of direct current bias $I_{dc}$ and $B$. AC excitation current used for this measurement is $I_{ac}$ = 2 nA. Fraunhofer-like oscillation patterns are clearly visible in (b). The inset in (b) shows $R_{xx}$ vs. $T$ for SC at optimal doping $\nu=2.10$. **(c)-(d)** Berezinskii–Kosterlitz–Thouless (BKT) measurements of differential resistance $dV_{xx}/dI$ versus $I_{dc}$ taken at different temperatures. BKT transition temperature $T_{BKT}$ is defined by fitting to $dV_{xx}/dI \propto I^2$. The insets show I-V curves that show critical current of approximately 0.69 µA ($\nu=-2.16$) and 0.06 µA ($\nu=2.10$). **(e)-(f)** Ginzburg-Landau coherence length measurements. Critical field $H_{c\perp}$ versus $T$ taken at half of normal state resistance values. The cyan lines are the best linear fit to data. We extract $\xi_{GL}$ = 38 nm (e) and $\xi_{GL}$ = 153 nm (f) from $H_c=(\Phi_0/(2\pi\xi_{GL}^2))(1-T/T_{c0})$, where $\Phi_0=h/2e$ is superconducting flux quantum and $T_{c0}$ is the mean-field temperature at zero $B$.

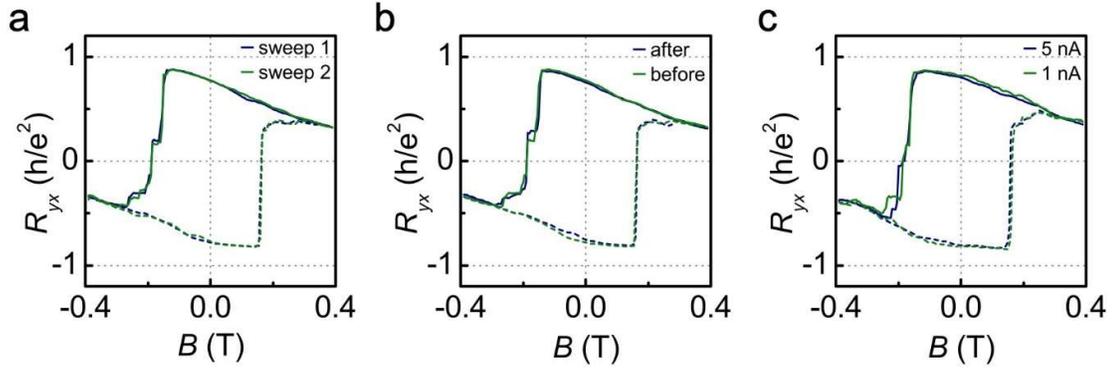

Extended Data Figure 10. | **Repeatability of hysteresis loops.** (a) $R_{yx}(B)$ line traces for two different sweeps taken at $\nu=+0.84$. The sweeps are taken 48 hours apart at 50 mK. (b) $R_{yx}(B)$ line traces taken before and after thermocycling the sample from 50 mK to 5 K and back to 50 mK at $\nu=+0.84$. (c) $R_{yx}(B)$ line traces taken for AC excitation currents 1 nA and 5 nA at $\nu=+0.82$. The dashed (solid) lines corresponds to ascending (descending) $B$-field.

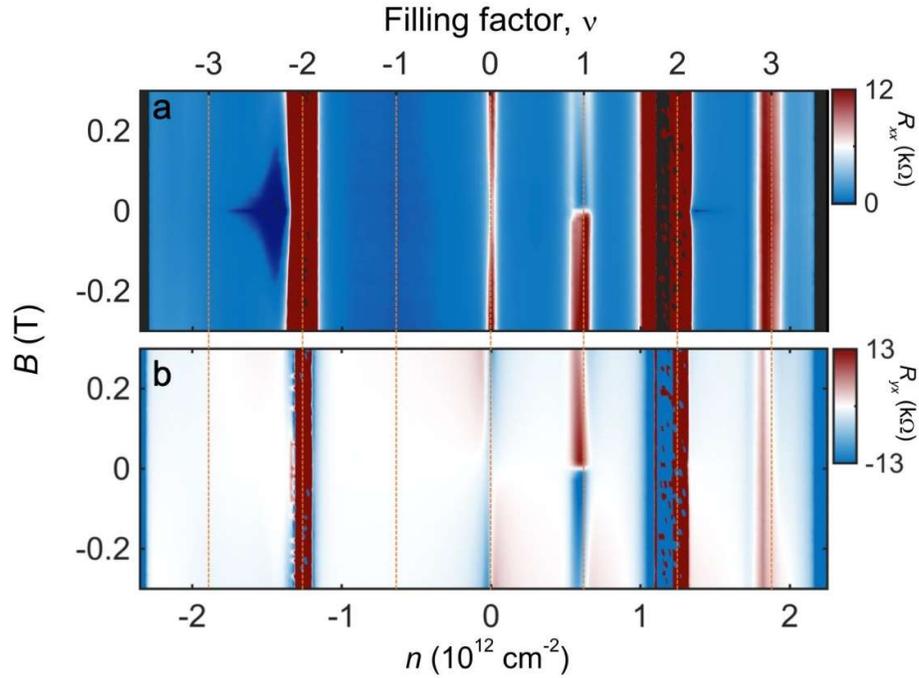

Extended Data Figure 11. | **High resolution scan for the low range of $B$-field.** (a) $R_{xx}$ and (b) $R_{yx}$ as a function of $n$ and $B$ for low range of the magnetic fields taken at 50 mK. Note an abrupt switch of sign of $R_{yx}$ at $\nu\approx+1$ upon crossing $B=0$ T line. Dark grey regions in (a) indicate quenched drain current due to the high sample resistance thus leading to unreliable voltage readings.

# G. Competition of correlated Chern insulator states under magnetic field at $\nu=+1$.

At integer fillings of the moiré flat bands, theoretical works have predicted a series of trivial and topological insulating states that are very close in energy. The ground state phase diagram seen by experiments vary from sample to sample which suggests that the exact energetic order of these states is sensitive to experimental parameters. When a magnetic field is applied, the total energies of these states as well as their relative order will be altered. Here we focus on the correlated insulating states at $\nu=+1$.

We perform self-consistent Hartree-Fock mean-field calculations to study the ground states at integer filling factors. Details of the calculation can be found in SI Ref[1] with an additional assumption that the remote band degrees of freedom are frozen, an approximation justified by the fact that the flat bands are relatively isolated from the remote bands. Here we used typical bandstructure parameters $\theta=1.1°$ and $T_{AA}/T_{AB}=0.6$, and, as an example, an interaction strength parameter $\epsilon^{-1}=0.03$ which can easily change as the distance to the nearby metallic gate is varied (see SI Ref[2]).

At $\nu=+1$, we found that the lowest energy states are insulating and spontaneously break $C_2T$ symmetry and the approximate U(4) flavor symmetry. As a result of flavor symmetry breaking, these states are both spin and valley polarized with one fully filled flavor, which is chosen spontaneously, and three half-filled flavors. The half-filled flavors have a gap between valence and conduction bands due to $C_2T$ symmetry breaking and a nonzero Chern number which, within the flat band Hilbert space, can be either +1 or -1. States with total Chern number $C=\pm 1$ and $\pm 3$ are thus possible with the former being slightly lower in energy as shown by Ref[3]. As discussed in the main text, at zero or low magnetic field one expects that $C=\pm 1$ states are more likely to be the ground state. The quasiparticle bands of the fully-filled flavor remains gapless with band touching at Dirac points. An overall energy gap is formed (orange region in Extended Data Figure 12) whose size depends on the strength of the Coulomb interaction. Several studies have also found ground states to be inter-valley coherent which we will discuss in later part of this section. In either case, the physics of the problem remains similar, as the charged ±1 excitation dispersions (Hartree-Fock bands) do not depend on the details of the polarization of the ground state (Ref.[4]).

The change in total energy due to applied magnetic field **B** can be described by the magnetic potential energy $\Delta E=-\mathbf{B}\cdot\mathbf{M}$, where **M** is the total magnetization, which includes both spin and orbital contributions. For orbital Chern insulators (see Ref[5]) considered here, they only differ in orbital magnetization so we can ignore the spin magnetization. We calculate the orbital magnetization approximately by treating the mean-field quasiparticle states as non-interacting Bloch electrons.

Extended Data Figure 12 shows the orbital magnetization of the $C=-1$ state which has one fully-filled flavor and three half-filled flavors. For the fully-filled flavor, Chern numbers of its conduction and valence bands cancel each other; because the fully-filled density matrix cannot have $C_2T$ breaking order within the flat-band Hilbert space, its magnetization vanishes. For half-filled flavors, the Chern numbers can be either +1 or -1 depending on their spontaneous sublattice polarization. Inside the gap, orbital magnetization is linearly proportional to the Chern number (e.g. Ref[5]) and its slope has the same sign as the Chern number.

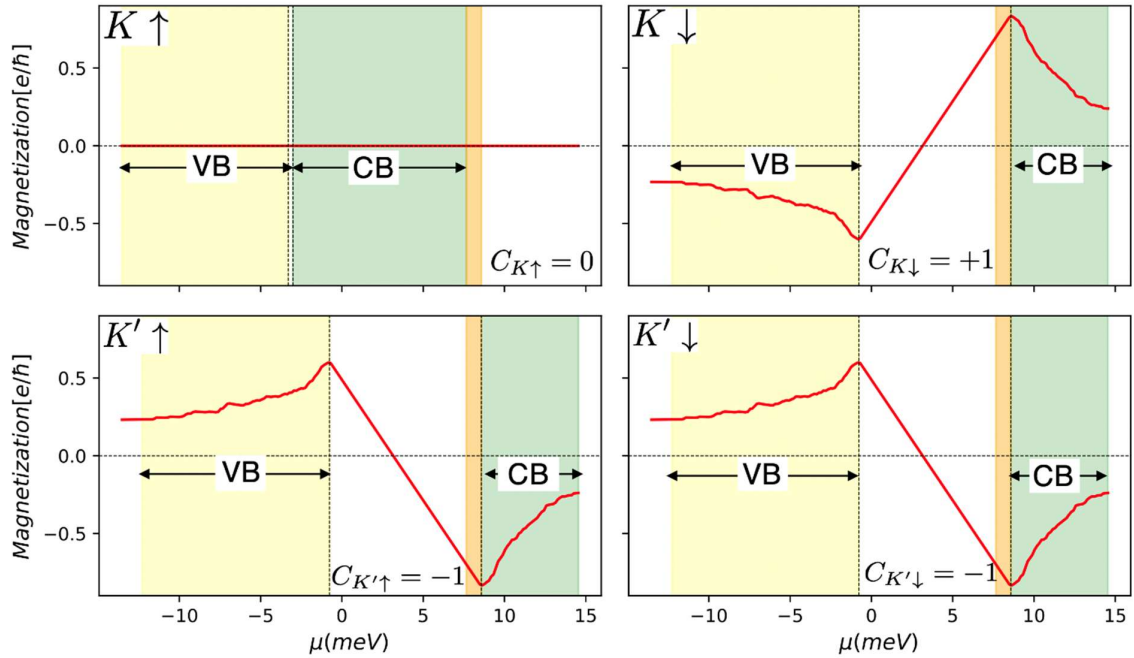

**Extended Data Figure 12. | Magnetization of the ($C$, $\nu$) = (-1, +1) mean-field insulating state as a function of chemical potential $\mu$.** The insulating state has one flavor fully filled flavor $K\uparrow$, which remain gapless, and three half-filled flavors $K\downarrow, K'\uparrow$, and $K'\downarrow$, all of which have an interaction induced gap. The yellow and light green regions mark the energy span of the quasiparticle valence band (VB) and conduction band (CB), respectively. The orange region marks the overall energy gap, once kinetic energy has been introduced (the non-flat-band case, where the kinetic energy in the BM model is also taken into account, along with the strong Coulomb interaction).

The total orbital magnetization **M** is a sum of the contributions from the four flavors. The sign of **M** inside the gap is in general not definite because the (overall) gap is very likely to span a larger energy range which includes the point at which **M** changes sign. For instance, at stronger interaction, the fully-filled flavor is lowered more in energy due to enhanced exchange splitting such that the overall gap is increased. If we consider the slightly doped cases on the hole or electron sides of $\nu$=+1, **M** has a definite sign. Because **M** normally changes sign across the gap, one expects the hysteresis loop in the Hall resistance also flips sign when going from hole doping side ($\nu$<+1) to electron doping side ($\nu$>+1). This property also agrees with the experimental observations in Figure 4d and 4e. The magnitude of the magnetization is not exactly the same on the two sides of the gap. When the non-local tunneling effect is included (see discussions in the next section), the magnitude of the magnetization is much smaller on the $\nu$>+1 side. The fact that the hysteresis loop changes sign at $\nu$=+1 suggests that $C$=+1 state is seen on $\nu$>+1 side, but only weakly developed. This agrees with the nonlocal tunneling effect that magnetization is small on the $\nu$>+1 side, so it is hard for magnetic field to coarsen domain configurations.

At higher enough magnetic field, the magnetization energy eventually wins over the zero-field total energy difference and favors state with larger magnitude of magnetization. The states with $|C|$=3 have similar quasiparticle dispersions as the $|C|$=1 states, however, because magnetization of the half-filled flavors have the same sign, they have larger magnitude of magnetization. Indeed the $C$=+3 Chern insulator state is seen here at higher field $B$>1 T and

$\nu>+1+\delta$ ($\delta>0$) with a well-quantized Hall conductance as shown in Figure 4c. We also expect $C=-3$ in the hole doped side $\nu<+1-\delta$, however, it is not seen here possibly because at the filling factor range $\nu<+1-\delta$, the ground state remains flavor symmetry unbroken leading to four-fold degenerate Landau levels as is indeed seen in this range. Exactly at which filling factor the flavor symmetry breaking order emerges depends primarily on the location of the van Hove singularity as discussed in Ref[6].

In above mean-field calculations, we have not allowed inter-valley coherence. Several theoretical works, including the work (Ref[3]) by one of the authors, have predicted that states with valley pseudo-spin component in the in-plane direction have lower energy than the spin-valley polarized states. However, because valley coherence mixes bands with the same Chern number, and the fact that the predicted valley anisotropy is rather weak, we expect the change in quasiparticle characters comparing to the spin-valley polarized states is only qualitative and small. This is in agreement with Ref[4] where, in the flat-band limit, the same excitation spectrum is found above any of the ground state, be they valley coherent or valley polarized. As a numerical test, we performed self-consistent mean-field calculations allowing inter-valley channels. The resulting dispersions are showing in Extended Data Figure 13. We find that the inter-valley coherent (IVC) state is slightly lower in energy. Comparing the dispersion of the IVC state with that of the polarized state, we see that its dispersion only changes slightly with a small splitting in the region where bands from the two valleys are degenerate in the spin-valley polarized state. We thus expect that the physics presented above using the spin-valley polarized states holds in IVC state case.

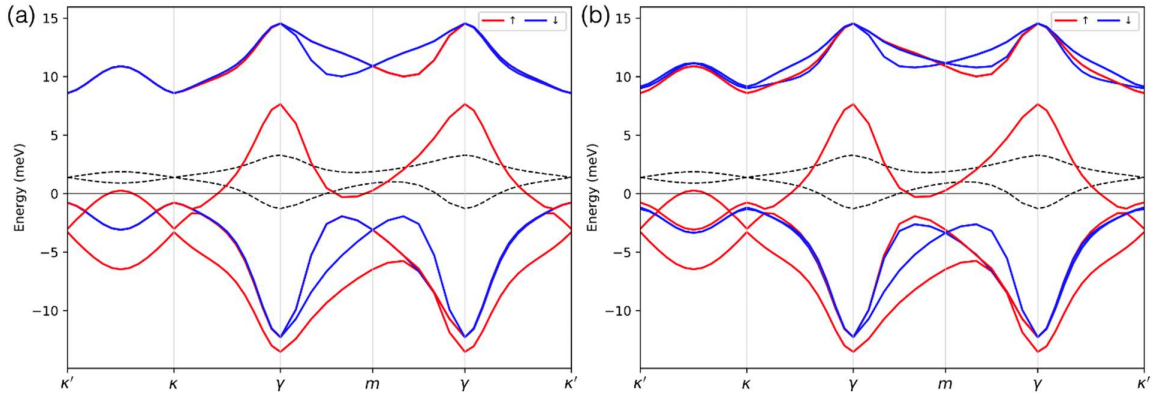

Extended Data Figure 13. | **Quasiparticle dispersion of (a) the spin-valley polarized state and (b) the inter-valley coherent state.** The red (blue) solid lines corresponds to spin up (down) bands. The spin up flavor has one more band filled than the spin down flavor. We note that along certain high symmetry lines the dispersions overlap. The dashed lines represent non-interacting bands.

A heuristic way of understanding the stabilization of the $C=-1$ state at low magnetic field (although during hysteresis, both $C=+1$ and $C=-1$ are observed and explained by our magnetization argument) is through their Landau level spectrum Extended Data Fig. 14a. Assume at zero magnetic field there exists two degenerate ground states, one with $C=+1$ and one with $C=-1$. These two ground-states have identical excitation spectrum. In field, they however differ by one zero mode, which, at constant filling $\nu=+1$, pins the Fermi level to be at the top/bottom of the valence/conduction band for Chern number +1/-1. By counting the energies of the occupied LL below the Fermi level, the $C=-1$ is favored in low B. The property that positive magnetic fields tend to favor $C=-1$ states for $\nu<+1$ and $C=+1$ states for $\nu>+1$ can be understood by the following heuristic argument. The total energy vs. filling factor curve has a cusp at the densities of the gaps illustrated in Extended Data Fig. 14b. For positive fields the $C=+1$ cusp occurs at a higher density than the $C=-1$ gap, leading to the illustrated energetic ordering illustrated vs. band filling.

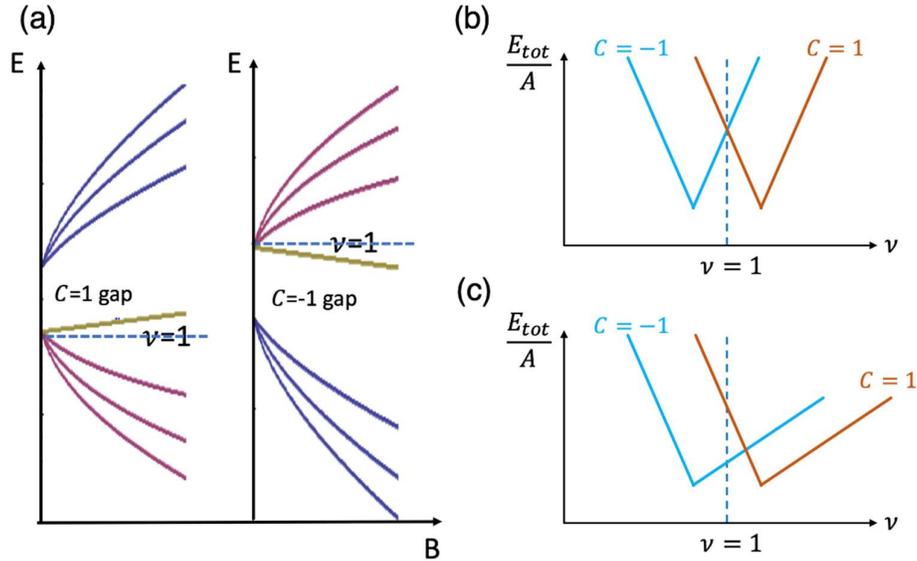

Extended Data Figure 14. | **(a)** Landau level spectrum of $C=+1$ and $C=-1$ Chern insulators with identical dispersion. The difference rests in the "zero-mode" Landau Level which has opposite in-field behavior for the two cases. **(b)** Total energy per area for states with $C=+1$ and $C=-1$ in the gap under finite positive magnetic field $B>0$. The cusps are where the $|C|=1$ gaps are located. **(c)** Same plot as in (b) for states with smaller magnetization on the electron doped side due to nonlocal tunneling effect discussed in next section.

## H. Effect of particle-hole asymmetry

The original Bistritzer-MacDonald (BM) continuum model has an approximate particle-hole symmetry based on which one would expect a particle-hole symmetric phase diagram with respect to $\nu=0$. Instead almost all the experiments have shown certain degree of particle-hole asymmetry in MATBG phase diagram. One possible origin, which is not due to extrinsic factors, is the nonlocal tunneling that is ignored in the BM model (see Ref[6] and references therein). Here we examine the effect of this nonlocal tunneling on the orbital magnetization.

To show the effect of the particle-hole asymmetry, we plot the orbital magnetization of non-interacting bands from a nonlocal continuum model. $\omega_{NL}$ characterizes the strength of the nonlocal tunneling which in turn determines the degree of particle-hole asymmetry. As $\omega_{NL}$ increases, the magnetization increases uniformly inside the gap and the point of zero magnetization shifts to the right side of the gap. The Chern bands used here all have $C=-1$ inside the gap. For bands with $C=+1$, the magnetization curve flips sign and decreases uniformly within the gap as $\omega_{NL}$ increases. For both cases, the point of zero magnetization shifts away from the middle of the gap towards the bottom of the conduction band. As discussed in the previous section, this effect causes the magnitude of the magnetization smaller on the high-filling factor side of the gap – in agreement with our experimental observations.

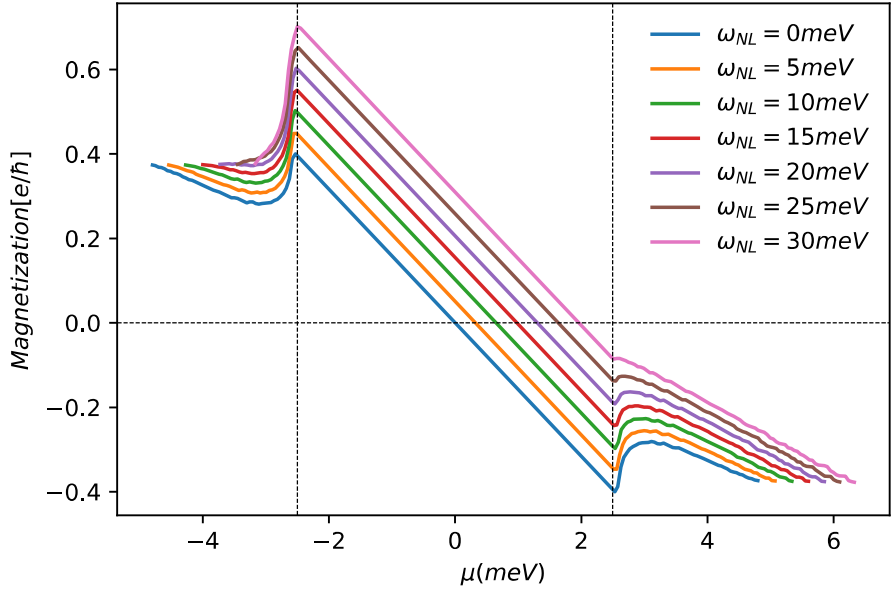

Extended Data Figure 15. | **Magnetization of particle-hole asymmetric flat bands.** The degree of the asymmetry is determined by the strength of the nonlocal tunneling $\omega_{NL}$. A $C_2T$ breaking potential is added manually which is necessary to generate finite orbital magnetization. The dashed lines on the left roughly marks the top of the valence band while that on the right roughly marks the bottom of the conduction band. We have shifted the energy axis, so the center of the gap for different $\omega_{NL}$ sit at same energy for the purpose of comparison.

**Supplementary References**